\documentclass[%
 twocolumn,
 reprint,
 superscriptaddress,
showpacs,
 amsmath,amssymb,
 aps,
prl,
]{revtex4}
\usepackage{color}
\usepackage{ulem}
\usepackage{graphicx}
\usepackage{dcolumn}
\usepackage{bm}

\def\rnum#1{\expandafter{%
\romannumeral #1}}
\def\Rnum#1{\uppercase\expandafter{%
\romannumeral #1}}

\begin{document}

\preprint{APS/123-QED\author{Takafumi Suzuki}}

\title{Unavoidable Gapless Boundary State and Boundary Superfluidity of 
Trapped Bose Mott States in Two-Dimensional Optical Lattices}

\author{Takafumi Suzuki}
\affiliation{Research Center for Nano-Micro Structure Science and Engineering, 
Graduate School of Engineering, University of Hyogo, Himeji, 
Hyogo 671-2280, Japan}
\author{Masahiro Sato}%
\affiliation{Department of Physics and Mathematics, Aoyama Gakuin University, 
Sagamihara, Kanagawa 252-5258, Japan}%
\affiliation{Advanced Science Research Center, Japan Atomic Energy Agency, Tokai 319-1195, Japan}
\affiliation{ERATO, Japan Science and Technology Agency, Sendai 980-8577, Japan}
\date{\today}

\begin{abstract}
We study the boundary nature of trapped bosonic Mott insulators 
in optical square lattices, by performing quantum Monte Carlo simulation.
We show that a finite superfluid density generally emerges in the incommensurate-filling 
(IC) boundary region around the bulk Mott state, 
irrespectively of the width of the IC region. 
Both off-diagonal and density correlation functions in the IC boundary 
region exhibit a nearly power-law decay.  
The power-law behavior and superfluidity are well developed 
below a characteristic temperature. 
These results indicate that a gapless boundary mode always emerges 
in any atomic Mott insulators on optical lattices. 
This further implies that if we consider a topological insulating state 
in Bose or Fermi atomic systems, 
its boundary possesses at least two gapless modes (or coupled modes) 
of an above IC edge state and the intrinsic topologically-protected edge state. 
\end{abstract}

\pacs{05.30.Jp,37.10.Jk,73.43.-f}
%
\maketitle

%
{\it Introduction}.$-$
Topological insulators (TIs)~\cite{review1,review2}, more widely, symmetry-protected-topological (SPT) states~\cite{SPT1,SPT2,Chen}, 
have been vividly studied as new quantum many-body states in the last decade. 
These gapful states cannot be characterized by any {\it local} order parameter, while they generally possess a gapless edge/surface mode. 
Each SPT phase is protected by certain symmetries, namely, it is stable against any perturbation keeping the symmetries. 
A complete classification of TIs and the relationship between their bulk symmetry and the corresponding surface/edge state have been established for free fermion models~\cite{Furusaki, Ryu, Kitaev}. 
Several TI materials have been synthesized and their surface/edge states have been observed~\cite{review1,review2}.

Many physicists stimulated by the study of fermionic TIs 
have been exploring SPT states in spin and boson systems. 
The Haldane-gap state~\cite{Haldane1,Tasaki} of one-dimensional (1D) spin-1 
antiferromagnets is a typical SPT state in quantum spin systems, 
and it is indeed realized in several quasi-1D magnets~\cite{Renard, Hagiwara}. 
In addition to the Haldane state, 
several SPT phases in 1D fermion, boson and spin systems have been discussed. 
In fact, a way to classify 1D bosonic SPT phases has been proposed 
by tensor product representation~\cite{Shuch}.

On the other hand, two- or three-dimensional (2D or 3D) bosonic SPT phases 
and their edge/surface states have been little understood. 
Several theorists discussed the possibility of higher-dimensional 
bosonic TIs~\cite{Kitaev2,Chen,Vishwanath, Metlitski,Ye,Liu}, and proposed 
ways to classify them: bosonic TIs in spatial dimensions 
$d$ can be distinguished by a technique based on $(d+1)$th group 
cohomology~\cite{Vishwanath, Metlitski,Ye,Liu}. 
In those studies, some models for bosonic TIs were predicted, 
but it is difficult to realize them in real materials because 
the corresponding Hamiltonians contain various tuned coupling constants.

For the realization of 2D or 3D SPT phases in boson or spin systems, 
strong interactions among bosonic particles or spins are generally necessary. 
The interaction usually makes it quite difficult to analyze the systems, 
and this is a main reason why the theory for 2D or 3D bosonic TIs 
has not been developed in comparison with fermionic TIs. 
Because of the same reason, even {\it non-topological} (i.e., trivial) 
gapped phases and their boundary nature have not been understood well 
in the strongly interacting boson and spin systems. 
In a sense, boundary nature of gapped states is more important 
than classification of ground states because physical phenomena 
at boundary can be observed and their information often provides 
a experimental way to characterize the bulk state.

Recently, we have studied a edge state of 2D spin-Peierls states~\cite{SS}, 
by quantum Monte Carlo (QMC) calculations~\cite{QMC1,QMC2,QMC3}. 
The Peierls state is a typical trivial gapped state in quantum spin systems 
and it does not accompany any spontaneous breaking of basic symmetries 
(such as spin rotation and time-reversal symmetries). 
We showed that if we prepare a sufficiently clean edge of the Peierls 
state with a large enough length ($\sim$ 50 sites), 
we can observe a gapless Tomonaga-Luttinger-liquid (TLL) like 
behavior~\cite{TL} along the edge and the edge spin-spin correlation 
function decays in an almost algebraic fashion. 
We proposed some experimental ways of 
detecting these gapless edge excitations.
In this paper, we will explore the fundamental nature of boundary states of 2D 
Bose Mott insulator on optical lattices, by QMC computations. 
Similarly to the spin-Peierls state and fermionic TIs, 
no spontaneous symmetry breaking occurs and a finite bulk excitation 
gap exists in Bose Mott states. 
They have been already realized in ultracold-atomic systems on optical 
lattice~\cite{Greiner,Trotzky,Zhang,Ha,Hung}. 
Therefore, their boundary nature could be 
an important research subject as a good comparison in that of bosonic TIs.

An important feature of trapped ultracold-atom systems is that 
their boundary is always clean. This considerably contrasts with 
solid systems, whose boundary is usually dirty. 
As a result, we always observe a clean and homogeneous boundary 
region in Bose Mott states. 
We will numerically clarify the boundary properties of 2D Bose Mott 
states. They could be experimentally detected in principle. 
Our findings are also useful to deeply understand boundary 
states of cold-atom TIs as well as those of Bose Mott states.

\begin{figure}
  \begin{center}
   \includegraphics[width=8cm]{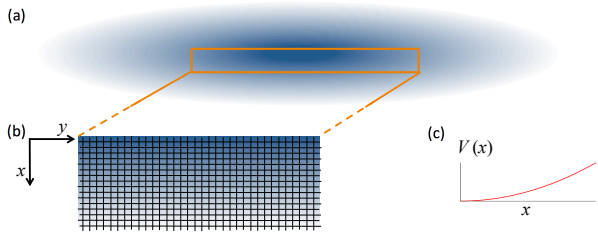}
  \end{center}
\caption{\label{fig1} (color online) 
(a) Full region of a 2D trapped Bose system on a square lattice. 
(b) Quasi-1D geometry we consider and (c) the corresponding 
confinement potential. 
Density of color stands for the depth of chemical potential. }
\end{figure}

{\it Model}.$-$ 
In this paper, we focus on the 2D soft-core Bose Hubbard model 
with confinement potentials. 
To discuss the finite-size effects systematically, 
we consider systems on the quasi-1D geometry shown in Fig.~\ref{fig1}(b).
This geometry would be hard to be realized in optical lattices, 
but it could be regarded as a boundary part of circular or elliptic 
shaped trapped systems [see Fig.~\ref{fig1}(a)].
The Hamiltonian for the quasi-1D geometry is given as 
\begin{eqnarray}
{\mathcal H}&=& -t\sum_{x,y} (b_{x, y}b^{\dagger}_{x+1,y} 
+b_{x, y}b^{\dagger}_{x,y+1} + h.c.) \nonumber\\
 & &+U\sum_{x,y}n_{x,y}^2 - \sum_{x,y}V(x)n_{x,y},
\label{Ham1}
\end{eqnarray}
where $V(x)={\mu_0 + \alpha x^2}$, $n_{x,y}=b^{\dagger}_{x,y}b_{x,y}$ and 
$b_{x,y}$ ($b^{\dagger}_{x,y}$) is a boson creation (annihilation) operator 
at position $(x,y)$. Parameters $t$, $U$, and $V(x)$ denote 
hopping amplitude, on-site repulsion, and 
axial confinement potential along the $x$-axis direction, respectively. 
To realize the quasi-1D geometry, we impose the periodic (open) boundary 
condition along the $y$ ($x$) direction. 
In our computations, we fix $U/t=20$, at which the bulk system can belong 
to the Mott-insulating state with a single boson per 
site~\cite{Greiner,Trotzky,Zhang,Ha,Hung}. 
We also fix the $x$-direction length $L_x=48$, but 
tune the $y$-direction length $L_y$. 
Increase of $\alpha$ means the growth of potential slope between bulk Mott 
and vacuum (empty) regions.

\begin{figure}[ht]
  \begin{center}
 \vspace{1cm}
   \includegraphics[width=0.9\linewidth]{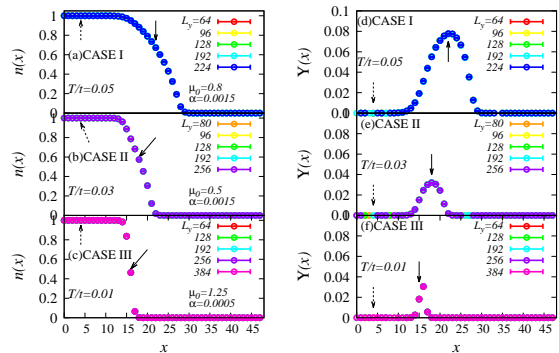}
\caption{\label{fig2} (color online) (a)-(c) Density and 
(d)-(f) stiffness ($\propto$ superfluid density) profiles estimated by QMC calculations
at extremely low temperatures $T$. 
Pairs [(a),(d)], [(b),(e)] and [(c),(f)] are the results of CASEs I, 
II and III of Table~\ref{list}, respectively. 
Solid (dotted) arrows denote positions of 
$x_0$ ($x_1$) in Table~\ref{list}. Results of systems with 
different sizes $L_y$ are almost degenerate.}
   \end{center}
\end{figure}
\begin{table}
\begin{tabular}{lcccccc}
 & $\mu_0/t$ & $\alpha/t$ & $x_0/L_x$ & $x_1/L_x$ & $\Delta x$ \\
\hline\hline
CASE I & 0.8 & 0.0015 & 0.458 & 0.0833  & 21 sites \\
CASE II & 0.5 & 0.0015 & 0.375 & 0.0833  & 10 sites \\
CASE III & 1.25 & 0.005  & 0.333 & 0.0833  & 4 sites \\
\hline \\
\end{tabular}
\caption{\label{list} Three parameter setups we chose, CASEs 
I, II, and III. The last three columns $x_0/L_x$, $x_1/L_x$ and $\Delta x$ 
denote representative positions in the incommensurate filling (IC) and 
the Mott regions, and the width of the IC region, respectively 
(see the text and Fig.~\ref{fig2}).}
\end{table}

{\it Boundary state in $T\to 0$ limit}.$-$ 
In order to understand the boundary nature of the Bose Mott state, 
we will study particle densities, superfluidity, and correlation functions 
of the model~(\ref{Ham1}) by QMC computations. 
In Fig.~\ref{fig2}, we first show the local density profile 
$n(x)=\langle n_{x, y}\rangle$ and 
local helicity modulus $Y(x)$, which is proportional to the local 
superfluid density~\cite{Fisher,note1}, changing $\mu_0$ and the 
curvature $\alpha$ at very low temperatures $T$ ($k_B$ is set to be unity). 
We here show QMC results of three parameter settings $(\mu_0,\alpha)$ 
in Table \ref{list} as representatives. 
In all CASEs I, II and III, an incommensurate (IC) filling region 
(e.g., $0.21<x/L_x<0.6$ in CASE I) appears between the filling-one Mott 
and the vacuum (empty) states. 
In the IC region, the local superfluid density takes a finite value. 
Size of the IC region and the superfluid density profile are 
almost irrespectively of the length $L_y$ when $L_y$ is sufficiently large. 
We note that finite-size effect along $x$ direction can be ignored 
in the present parameter settings. 
Survival of a superfluid density sharply contrasts with the case of 
the purely 1D Bose system because the latter superfluid density 
is known to disappear in the thermodynamic 
limit~\cite{Affleck,Oshikawa,note2}.  
We confirmed that a further narrower 
IC region still survives when a more large value of $\alpha$ is applied. 
Furthermore, we observe an IC region when we apply other types of 
the chemical potentials with non-harmonic 
curvatures~\cite{Gaunt13,Liang12}: 
$V(x)=\mu_1+\alpha_1 x^{10}$ and $=\mu_2+\alpha_2\exp(-x/\xi_2)$. 
These results indicate that an IC superfluid region between Mott and 
vacuum areas generally appears and a special potential $V(x)$ 
with an extremely large curvature is necessary to remove the IC region.

\begin{figure}[thb]
  \begin{center}
  \vspace{0.5cm}
   \includegraphics[width=0.9\linewidth]{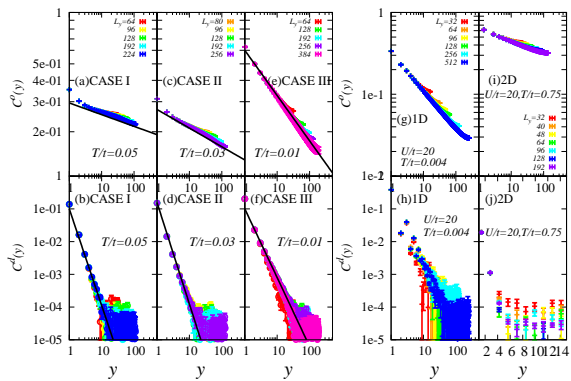}
\caption{\label{fig3}(Color online) Off-diagonal and density correlation 
functions along the $y$ direction. 
Pairs [(a),(b)], [(c),(d)] and [(e),(f)] are the results 
of CASE I, II, and III at the IC position $x=x_0$, 
respectively. Pairs [(g),(h)] and [(i),(j)] are respectively 
the results of spatially uniform 1D and 2D Bose Hubbard models 
at an IC filling. Black lines are guides to eyes.}
  \end{center}
\end{figure}

In Fig.~\ref{fig3}, we present the equal-time one-particle (off-diagonal) 
correlation function $C^o(y)$=$\langle b^{\dagger}_{x_0,y} b_{x_0,0} \rangle$ 
and density one $C^d(y)$=$\langle n_{x_0,y} n_{x_0,0} \rangle
-\langle n_{x_0,y}\rangle^2$ 
at the boundary position $x=x_0$ where $\rho_s(x_0)$ 
takes the largest values. 
As a comparison, we also show the QMC results of a purely 1D Bose Hubbard model [Fig.~\ref{fig3}(g) and (h)] and 
a spatially uniform 2D Bose Hubbard one [Fig.~\ref{fig3}(i) and (j)] in an IC-filling case. 
The 1D and 2D Bose systems of Fig.~\ref{fig3} belong to TLL and Kosterlitz-Thouless (KT) phases, respectively. 
At the position $x_0$, power-law decays along the $y$ direction are observed in both off-diagonal and density correlations with long distances. 
Their critical exponents are evaluated by assuming the form $C^z(y)$=$const. \times y^{{-\eta}_z}$, 
and they are summarized in Table \ref{expo}. 
\begin{table}
\begin{tabular}{lcccc}
 & CASE I & CASE II & CASE III & 1D \\
\hline\hline
$\eta_o$  & 0.069(2) & 0.12(4) & 0.25(1) & 0.443(8) \\
$\eta_d$  & 2.9(1)  & 3.1(2)  &  2.1(1) & 2.23(6) \\ 
\hline
\end{tabular}
\caption{\label{expo} Estimated decay exponents for off-diagonal and 
density correlations, assuming $C^z(y)\sim const. \times y^{-\eta_z}$, 
where $z=o$ ($d$) stands for the off-diagonal (density) correlation.}
\label{table2}
\end{table}
The emergence of algebraic decay is independent of the width of IC region. 
This clearly indicates that at least one gapless edge mode around 
the bulk Mott-insulating region always appears at very low temperatures. 
The algebraic decay of the density correlation is quite different from 
that of the KT phase in uniform 2D Bose systems. In fact, 
Fig.~\ref{fig3}(j) shows that the density correlation decays 
exponentially in the 2D case. In addition, it is known~\cite{TL} that 
two critical exponents satisfy $\eta_o\eta_d=1$ 
in the purely 1D TLL (see Table~\ref{table2}), while 
the relation is clearly broken in the present boundary gapless mode. 
These results conclude that the boundary IC region possesses 
intermediate properties between 1D and 2D Bose systems. 

\begin{figure}[htb]
  \begin{center}
   \includegraphics[width=0.8\linewidth]{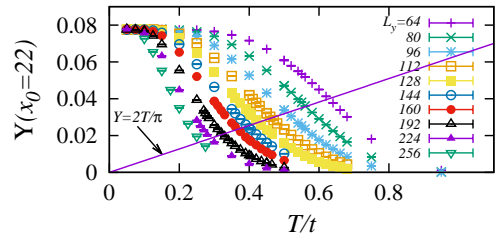}
  \end{center}
\vspace{-0.5cm}
\caption{\label{fig4} (color online) Local helicity modulus 
at a IC position $x=x_0$ in CASE I.}
\end{figure}

{\it Boundary state in finite temperatures}.$-$ 
Next, we discuss the temperature dependence of the boundary IC states. 
In the present model as well as real experimental systems, 
effects of finite size and spacial inhomogeneity may spoil 
true phase-transition phenomena in the thermodynamic limit, 
but their residual things might still survive. 
Figure \ref{fig4} shows the temperature dependence of local helicity 
modulus $Y(x_0)$ at a typical IC position $x=x_0$ in CASE I. 
The figure shows that the $L_y$ dependence of $Y(x_0)$ 
becomes negligibly small below $T/t \sim 0.075$,
where off-diagonal and density correlation 
functions decay algebraically. This small $L_y$ dependence 
implies the existence of a gapless KT-like phase 
around the bulk Mott state.

\begin{figure}
  \begin{center}
 \includegraphics[width=0.8\linewidth]{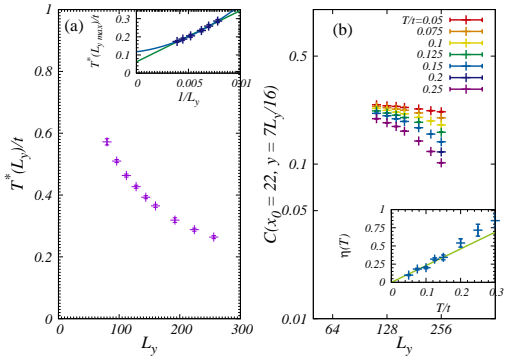}
  \end{center}
\vspace{-0.5cm}
\caption{\label{fig5} (a) Size dependence of characteristic temperature 
$T^{*}(L_y)$. (b) Temperature dependence of the exponent $\eta_o$.}
\end{figure}

To quantitatively determine a KT-transition-like temperature 
in our finite inhomogeneous system, we simply apply the standard finite-size 
analysis for the KT transition in spatially uniform 2D systems, which 
is expected to be reliable if the width of the IC region is sufficiently large 
as in CASE I. In the uniform system, $Y$ approaches to $2 k_B T_{KT}/ \pi$ 
at the KT transition temperature $T=T_{KT}$. 
For each finite system with size $L$, 
the KT transition temperature $T^*(L)$ is known to be 
\begin{eqnarray}
T^*(L)=T_{KT}(L \rightarrow \infty)\Big(1+\frac{1}{2 \ln L +C}\Big),
\label{eq:T_KT}
\end{eqnarray} 
where $C$ is a fitting parameter. 
Since we have the data of $Y(L_y)$ for different sizes $L_y$ 
in Fig.~\ref{fig4}, 
the temperature $T^*(L_y)$ can be estimated as the cross point 
between numerically determined $Y(L_y)$ in Fig.~\ref{fig4} and 
the linear line $Y(T)=2 k_BT/ \pi$. 
In Fig.~\ref{fig5}(a), we plot the evaluated $T^*(L_y)$. 
In its inset, we determine the value 
of $T^*(L_y \rightarrow \infty)$ by combining $T^*(L_y)$
and the scaling relation~(\ref{eq:T_KT}) ($T^*(L_y \rightarrow \infty)$ 
corresponds to the KT transition temperature $T_{KT}$ 
in the case of uniform systems). 
The characteristic temperature $T^{*}(\infty)$ is determined as 
$T^{*}(\infty)/t \sim 0.11(1)$.

As an alternating way to determine $T^{*}(\infty)$, we can utilize 
correlation functions in our inhomogeneous system. 
It is well known that the critical exponent 
$\eta_o$ of the off-diagonal correlator increases from zero and 
it becomes a quarter at the KT transition with the growth of temperature. 
Let us simply apply this property to fix the KT-like temperature 
in our IC region. 
The inset of Fig.~\ref{fig5}(b) is the $T$ dependence of 
the exponent $\eta_o$.  
We see that $\eta_o$ indeed crosses a quarter around 
$T=T^{*}(\infty)$ which was estimated above from $Y(L_y)$.

We stress that the temperature $T^*(\infty)$ is much lower than the 
true KT transition temperature $T_{KT}$ in the uniform 2D system. 
From the QMC calculation, we obtained 
$T_{KT}/t \sim 0.92(3)$ for $U/t=20$ and $\mu/t=-0.74$,
where the averaged particle number per site is almost same as that 
at the position $x=x_0$ in CASE I. 
This must be because the development of off-diagonal correlation 
along the $x$ direction is suppressed 
owing to the existence of Mott-insulating and vacuum regions. 
When the width of the IC region is small as in CASE III, 
it is hard to quantitatively determine $T^*$. However, 
even in such a case, a KT-like power law in correlations appears 
at very low temperatures (see Fig.~\ref{fig3}).

%
%

\begin{figure}[htb]
\begin{center}
 \includegraphics[width=0.8\linewidth]{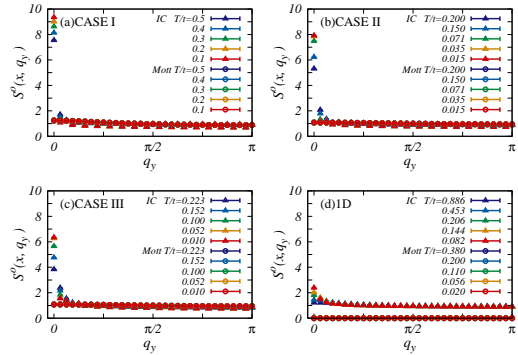}
\caption{\label{fig6} (color online) (a)-(c) $T$ dependence of 
structure factors $S^o(x,q_y)$ at $L_y$=$64$. 
Circles (triangles) are the results for the IC (Mott) region 
at $x=x_0$ ($x=x_1$). (d) $T$ dependence of the structure factor of 
the off-diagonal correlator $S^o(q)$ 
in a TLL state of 1D Bose Hubbard model. 
}
  \end{center}
\end{figure}

{\it Structure factors}.$-$ 
From all the discussions above, we see that at least a 
gapless IC state always appears around the Bose Mott state 
if temperature is low enough. 
Finally we discuss a experimental way to detect the gapless edge mode. 
In cold-atomic systems, the momentum distribution of correlation 
functions~\cite{Ueda-Pitaevskii} can be observed in principle. 
For example, time-of-flight (TOF) method and light-scattering 
spectroscopy have been applied to observe them. 
In Fig.~\ref{fig6}, we show the momentum-$q_y$ distribution of 
$S^o(x,q_y)=1/\sqrt{L_y}\sum_{y} C^o(x,y)e^{-i yq_y}$
for the IC region at $x=x_0$ and for the bulk Mott region 
at $x=x_1$. Here, $C^o(x,y)=\langle b_{x,y}^\dag b_{x,0}\rangle$. 
In the realistic experimental setup, 
the number of sites along the $y$-axis is less than $\sim$100 sites and 
then we set $L_y=64$ in all the panels of Fig.~\ref{fig6}. 
As temperature decreases, the momentum distribution 
at zero wave number $q_y=0$ of the IC region well develops,
while that of the Mott region is suppressed for 
any wave number $q_y$. 
The $q_y=0$ peak reflects the development of superfluidity 
in the IC region. 
In Fig.~\ref{fig6} (d), as a comparison to the IC region, 
we show the momentum distribution of 
a finite-size 1D Bose Hubbard model under an uniform chemical potential 
with the almost same filling as the IC boundary region. 
In the 1D case, we also observe a $p_y=0$ peak structure which is 
the contribution of TLL. 
From Fig.~\ref{fig6}, we find that momentum distributions in 
both Mott and IC regions exhibit the similar $T$ dependence 
to the finite-size 1D system with the same filling. 
This is an another evidence for the existence of a gapless edge mode 
in the IC region and it also indicates the difficulty of 
distinguishing the IC gapless state and the 1D TLL.

{\it Summary and discussions}.$-$
In conclusion, we have studied the edge state surrounding 
the 2D Bose Mott-insulating phase. From the QMC method, 
we have found that in the IC edge region, both off-diagonal and 
density correlators show an algebraic decay and 
a superfluid density appears below a characteristic temperature 
irrespective of the width of the IC region. 
This "universal" gapless edge mode can be detected e.g., 
by observing a $q_y=0$ peak of $S^o(x,q_y)$.

\begin{figure}[htb]
\begin{center}
 \includegraphics[width=0.8\linewidth]{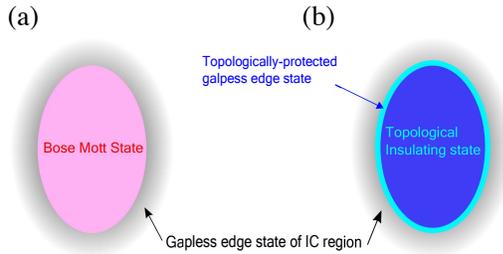}
\caption{\label{fig7} (color online) Spatial structures of a trapped 
Bose Mott state (a) and a trapped topological insulating state (b) 
in 2D cold atoms.}
  \end{center}
\end{figure}

Our result naturally indicates that a similar gapless edge mode generally 
emerges in any kinds of 2D cold-atomic Bose and Fermi insulating states. 
Therefore, if we consider a topological insulating state in 2D cold-atom 
systems, we can expect that there are at least two gapless modes of 
the edge state in the IC region and an intrinsic 
topologically-protected edge state as shown in Fig.~\ref{fig7}. 
There might be a relevant coupling between these two edge 
states. Thus, when we discuss a way to detect topological 
edge mode in 2D cold-atom systems, we should generally consider 
effects of a non-topological (but universal) edge mode in the IC region 
around the bulk insulating area. 

{\it Acknowledgment}.$-$ 
We would like to thank Yoshiro Takahashi and Akiyuki Tokuno for providing us some information of recent techniques of trap potentials.
This research partially used computational resources of the K computer provided by the RIKEN Advanced Institute for Computational Science through the HPCI System Research project (Project ID:hp130081).
We also thank numerical resources in the ISSP Supercomputer Center of the University of Tokyo and cluster machines in Nano-micro structure science and engineering, University of Hyogo. 
This work was financially supported by KAKENHI (Grants No. 25287088, 25287104, 26870559).


\begin{thebibliography}{99}
\bibitem{note2}
We note that even in 1D systems, a non-zero superfluid density 
can survive if the system size is finite. See Refs.~\cite{Affleck,Oshikawa}.
\end{thebibliography}

\end{document}



\title{Supplemental Material: Local Superfluid Density}

\author{Takafumi Suzuki}
\affiliation{Research Center for Nano-Micro Structure Science and Engineering, 
Graduate School of Engineering, University of Hyogo, Himeji, 
Hyogo 671-2280, Japan}
\author{Masahiro Sato}%
\affiliation{Department of Physics and Mathematics, Aoyama Gakuin University, 
Sagamihara, Kanagawa 252-5258, Japan}%
\affiliation{Advanced Science Research Center, Japan Atomic Energy Agency, Tokai 319-1195, Japan}
\affiliation{ERATO, Japan Science and Technology Agency, Sendai 980-8577, Japan}



\maketitle


In Supplemental Material, we define and explain 
local helicity modulus and local superfluid density. 
The contents mainly focus on the numerical techniques of 
quantum Monte Carlo (QMC) simulation. 
Readers who are interested in only physical results may skip 
this Supplement Material.

The helicity modulus $Y_{\rm tot}$ 
and the superfluid density $\rho_{s}^{\rm HOMO}$ 
in a $d$ dimensional $uniform$ system~\cite{Fisher}
are defined by using a winding number vector 
${\boldsymbol W}$~\cite{Polleck} as 
\begin{eqnarray}
Y_{\rm tot}=t d\rho_{s}^{\rm HOMO}=L^{2-d} 
\frac{\langle {\boldsymbol W}^2 \rangle}{\beta},
\label{eq1}
\end{eqnarray}
where $t$ is the "averaged" short-range hopping amplitude, 
$L$ is a linear size of the system.
The $\alpha$-component $W_\alpha$ of ${\boldsymbol W}$ stands for the number of winding particles 
along the $\alpha$ axis in the path-integral formalism, and can be evaluated by QMC method. 
The value $Y_{\rm tot}$ is usually insensitive to details of the system, 
but $\rho_s^{\rm HOMO}$ may depend strongly on them such as values of 
hopping amplitudes and interactions. In fact, for instance, it is well known 
that the helicity modulus exhibits a universal jump at a Kosterlitz-Thouless 
(KT) transition point in two dimensions.

Now, keeping in mind the above facts of the uniform systems, 
let us discuss local superfluidity and local helicity modulus 
in spatially inhomogeneous systems. 
To see the universal nature of the systems, we will mainly 
use the local helicity modulus rather than local superfluid density. 
In order to theoretically define helicity modulus, 
we have to impose a periodic boundary condition 
along at least one spatial direction, because the modulus is 
defined by the energy variation after slightly twisting 
the boundary condition. 
Namely, by definition, it is impossible to define "truly" local helicity 
modulus at any position ${\boldsymbol x}$ on the $d$ dimensional space. 
However, when $d\geq 2$, we can introduce a local helicity modulus 
defined at any position on one special $x$ direction by 
imposing a periodic boundary condition along a direction perpendicular 
to the $x$ axis. 
For a $d$-dimensional system with $d\geq 2$, 
let us define the local helicity modulus $Y(x)$ on $x$ axis 
such that the total helicity modulus is given by summation of $Y(x)$ 
along the $x$ direction:
\begin{eqnarray} 
Y_{\rm tot}=\sum_{x}Y(x). 
\label{eq2}
\end{eqnarray} 
Following Eqs.~(\ref{eq1}) and (\ref{eq2}), we then define 
the relation between the local superfulid density $\rho_s(x)$ 
and $Y(x)$ as 
\begin{eqnarray}
\rho_s(x)=\frac{Y(x)}{t d}.
\end{eqnarray}
In this equation, $\rho_s(x)$ stands for the superfluid 
density along a direction perpendicular to the $x$ axis. 
For example, if we consider a two dimensional system with 
open boundary condition along the $x$ axis, 
$\rho_s(x)$ is the superfluid density 
along the $y$ direction at $x$.

Next we explain how the local helicity modulus is related to 
numerical evaluated quantities such as ${\boldsymbol W}$.
Similarly to local helicity modulus, winding (topological) numbers 
including ${\boldsymbol W}$ also has a non-local nature 
and a periodic boundary condition is necessary 
along at least one direction to define them. 
Paying attention to these properties, we can introduce 
the partial winding number vector ${\boldsymbol w}(x)$ 
at each position on the $x$ axis. 
The summation of ${\boldsymbol w}(x)$ over the $x$ axis is 
equivalent to the original winding-number vector ${\boldsymbol W}$: 
\begin{eqnarray}
{\boldsymbol W}=\sum_{x}{\boldsymbol w}(x). 
\label{eq3}
\end{eqnarray}
The $x$ component $w_x(x)$ of ${\boldsymbol w}(x)$, which can be evaluated 
by QMC, does not satisfy the property of winding number, since 
information about the whole range of the $x$ axis is necessary 
to define winding number along the $x$ axis. However, 
sum of them, i.e., $W_x$ can be a winding number if we impose 
a periodic boundary condition along the $x$ direction. 
Each remaining component ${w_\alpha}(x)$ ($\alpha \neq x$) is the number of winding particles 
along the $\alpha$ direction at $x$, and thus it has a topological nature.

From Eqs.~(\ref{eq1}), (\ref{eq2}) and (\ref{eq3}), 
we find that ${\boldsymbol w}(x)$ and $\rho_{s}(x)$ 
should satisfy the following relation,
\begin{eqnarray}
Y(x) = L^{2-d} 
\frac{\langle {\boldsymbol W}\cdot {\boldsymbol w}(x) \rangle }{\beta}.
\end{eqnarray}
When an open boundary condition is applied along the $x$ axis 
as in our setup in the main text, 
the $x$ component of the winding vector becomes zero and 
$\rho_s(x)$ represents the superfluid density along the $y$ axis 
at the position $x$. 
In this way, we can evaluate the local helicity modulus and 
local superfluid density (see Fig.2 of the main text).
